\documentclass[prl,aps,floatfix,twocolumn]{revtex4}
\usepackage{endnotes}
\usepackage{latexsym,amssymb}
\usepackage{endnotes}
\usepackage{latexsym,amssymb}
\usepackage{amsmath, amsthm, amssymb}
\usepackage{graphicx}% Include figure files
\usepackage{epsfig}
\usepackage{subfigure}

\begin{document}
\title{Self Assembly of Janus Ellipsoids}
\author{Ya Liu}
\affiliation{Department of Physics, Lehigh University, Bethlehem, PA 18015}
\author{Wei Li}
\affiliation{Department of Physics, Lehigh University, Bethlehem, PA 18015}
\author{Toni Perez}
\affiliation{Department of Physics, Lehigh University, Bethlehem, PA 18015}
\author{J. D. Gunton}
\affiliation{Department of Physics, Lehigh University, Bethlehem, PA 18015}
\author{Genevieve Brett}
\affiliation{Department of Physics, Skidmore College, Saratoga Springs, NY 12866}
%\pacs{PACES numbers: 82.35.Pq Biopolymers, biopolymerization, 87.15.-v Biomolecules: structure and physical properties
% 83.80.Rs Polymer solutions rheology of, 05.40.-a Random processes, }

\begin{abstract}

We propose a primitive model of Janus ellipsoids that represent particles with an ellipsoidal core and  two  semi-surfaces coded with dissimilar properties, for example, hydrophobicity and hydrophilicity, respectively. We investigate the effects of  the aspect ratio on the self-assembly morphology and dynamical aggregation processes using Monte Carlo simulations. We also discuss certain differences between our results and those of earlier results for Janus spheres. In particular, we find that the size and structure of the aggregate can be controlled by the aspect ratio.

\end{abstract}

\maketitle
\section{Introduction}
Colloidal particles with anisotropic properties interact through an energy that depends not only on their spatial separation but also on their relative orientations.  This is a relatively new field that  has been receiving considerable attention in recent literature \cite{Glotzer2007, Kegel2011, QianChen2011a, Pawar2009, QianChen2011, Romano2011}. Various site-specific techniques such as template-assisted fabrication and physical vapor deposition have been developed to synthesize patchy colloidal particles of different shapes, patterns and functionalities \cite{Pawar2010, Jiang2010, Manoharan2003}. The self-assembly of those building blocks into a desired mesoscopic structure and function are considered as a bottom-up strategy to obtain new bulk materials that have potential applications in broad fields including  drug delivery, photonic crystals, biomaterials and electronics. The effects of anisotropy have been classified by Glotzer and Solomon using the concept of an anisotropy dimension including patchiness, aspect ratio, faceting etc \cite{Glotzer2007, Zhang2004}. 

Patchiness on spheres with different number, size and arrangements have been successfully fabricated in colloids experiments and reveal interesting properties and crystal structures \cite{QianChen2011a, QianChen2011, Pawar2009, Hong2006}. One particular example is the Janus sphere with two dissimilar semi-surfaces,  which has been extensively studied by experiments and theories \cite{Sciortino2009, Miller2009, QianChen2011a, Sciortino2010}.  Examples of dissimilar, coded surfaces include hydrophobic/hydrophilic,  charged/uncharged, and metallic/polymer surfaces. A variety of stable structures and unusual phase behaviors have been found under different chemical conditions of the solution by both experiments and computer simulations. Without losing the generality of the dissimilarity of two surfaces, a primitive two-patch Kern-Frenkel model has been used to model Janus spheres \cite{Kern2003, Sciortino2009}. In Monte Carlo simulations, this simplified model reproduces the main experimental features including self-assembly morphology and sheds light on potential applications in engineering and theoretical studies of reentrant phase diagrams \cite{Sciortino2010, Reinhardt2011}. 

As suggested by Glotzer and Solomon, it's natural to extend the study to explore the role of aspect ratio, in which patches are arranged on an anisotropic core such as spheroids. The anisotropy dimension is related to the aspect ratio and recent studies using ground-state energy calculation reveal many interesting structures such as helix \cite{Fejer2007, Fejer2011}.  However,  how the aspect ratio affects the self-assembly morphology is still not clear. In experiments,  ellipsoidal colloids can be engineered with high monodispersity using techniques such as deforming the spherical silica by ion fluence \cite{Dillen2004, Ho1993}, which makes the patched ellipsoidal surface possible if one can combine this with the template-assisted fabrication technique. Recently, unpublished work shows that Janus football-like ellipsoid has been fabricated \cite{DeConinck}.  Suspensions of Janus ellipsoidal particles provide a good candidate to study the effect of aspect ratio on self-assembly and new feature of colloidal phase transformation. 

In this report, we first propose a theoretical model of  Janus ellipsoids with hard-core repulsion and quasi-square-well attraction. The properties of the model are also discussed.  In the second part, we use Monte Carlo simulations to study the effect of aspect ratio of ellipsoids on the self-assembly morphology and dynamical properties. In the last section we present a brief conclusion. 

\section{Model}
In our simulation study, a primitive model is presented for Janus spheroidal particles, with the lengths of  the principle axes denoted by $a \neq b = c$. Depending on the aspect ratio, defined as $\epsilon = \frac{a}{b}$, the ellipsoid is charaterized as oblate if $\epsilon < 1$ ("M $\&$ M") and prolate if $\epsilon > 1$ ("football").  Consider an ellipsoid centered at $\mathbf{r}_0 = (x_0, y_0, z_0)$ whose axial orientations are given by
$\mathbf{u} ^T= ( u_1, u_2 ,u_3)$, $\mathbf{v}^T= (v_1,v_2,v_3)$, $\mathbf{w}^T = (w_1,w_2,w_3)$.  The equation for this ellipsoid has the explicit form
\begin{eqnarray}
\frac{[\mathbf{u}\cdot (\mathbf{r-r_0})]^2}{a^2} + \frac{[\mathbf{v}\cdot (\mathbf{r-r_0})]^2 + [\mathbf{w}\cdot (\mathbf{r-r_0})]^2}{b^2} =1\nonumber\\
 \label{eq:generalellipsoid}
\end{eqnarray}
In terms of a matrix representation, Eq. \ref{eq:generalellipsoid}  can be rewritten as $\mathbf{X}^T\mathbf{A}\mathbf{X} = 0$, where $\mathbf{X}^T = (x,y,z,1)$ and $\mathbf{A}$ is a $4\times 4$ symmetric matrix. The ten independent elements of $\mathbf{A}$ are listed as follows:\\
1) $\mathbf{A}_{11} = \frac{u_1^2}{a^2} + \frac{v_1^2}{b^2} + \frac{w_1^2}{c^2} $ , $\mathbf{A}_{12} = \frac{u_1u_2}{a^2} + \frac{v_1v_2}{b^2} + \frac{w_1w_2}{c^2}
$\\
  $\mathbf{A}_{13} = \frac{u_1u_3}{a^2} + \frac{v_1v_3}{b^2} + \frac{w_12_3}{c^2} $ , $\mathbf{A}_{14} = -x_0\mathbf{A}_{11} -y_0\mathbf{A}_{12}  -z_0\mathbf{A}_{13}  $\\
2)  $\mathbf{A}_{22} = \frac{u_2^2}{a^2} + \frac{v_2^2}{b^2} + \frac{w_2^2}{c^2} $ , $\mathbf{A}_{23} = \frac{u_2u_3}{a^2} + \frac{v_2v_3}{b^2} + \frac{w_2w_3}{c^2}
$\\
$\mathbf{A}_{24} = -x_0\mathbf{A}_{21} -y_0\mathbf{A}_{22}  -z_0\mathbf{A}_{23}  $\\
3)  $\mathbf{A}_{33} = \frac{u_3^2}{a^2} + \frac{v_3^2}{b^2} + \frac{w_3^2}{c^2} $ , $\mathbf{A}_{34} = -x_0\mathbf{A}_{13} -y_0\mathbf{A}_{23}  -z_0\mathbf{A}_{33}  $\\
$\mathbf{A}_{44} = -1 -x_0\mathbf{A}_{14} -y_0\mathbf{A}_{24}  -z_0\mathbf{A}_{34} $.\\  
Here $\mathbf{A}$ is  normalized so that a point in the interior of the ellipsoid $\mathbf{X}_0$ satisfies $\mathbf{X}_0^T\mathbf{A}\mathbf{X}_0 < 0$ and thus $\det(\mathbf{A}) < 0$.  This normalization will be used to determine the spatial relation between ellipsoids in the late section.

\begin{figure}[htbp]
\centering
\includegraphics[width=0.85\linewidth]{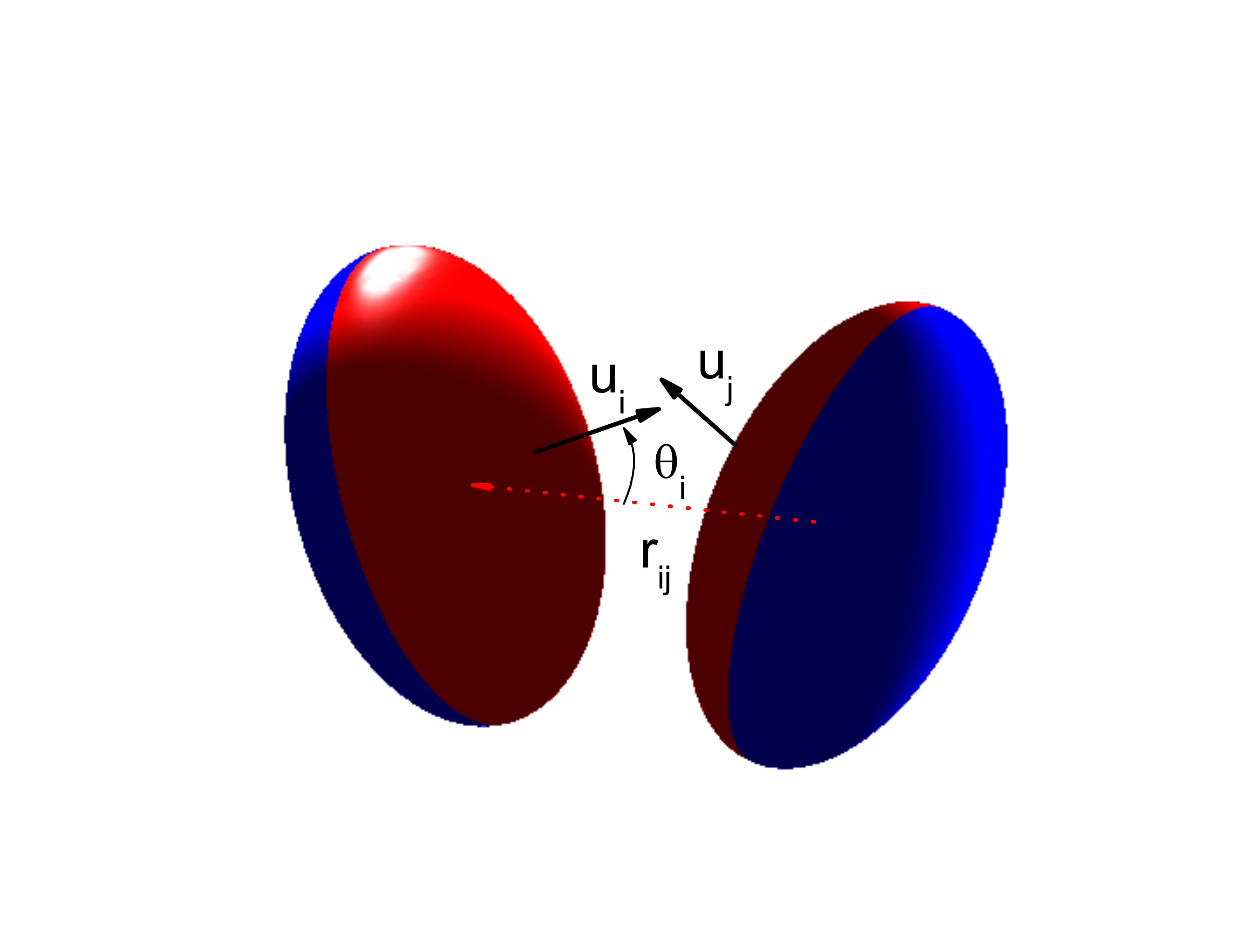}
\caption{ Plot of two interacting ellipsoids labeled as i and j, in which attractive and hardcore repulsive surfaces are coded by red and blue, respectively. $\mathbf{r}_{ij}$ is center-to-center displacement pointing from j to i. $\mathbf{u}_i$ and $\mathbf{u}_j$ is the patchy orientation. $\theta_{i}$ is the patch angle of ith ellipsoid. }
\label{fig:janusellipsoid}
\end{figure}

\begin{figure}[htbp]
\centering
\includegraphics[width=0.95\linewidth]{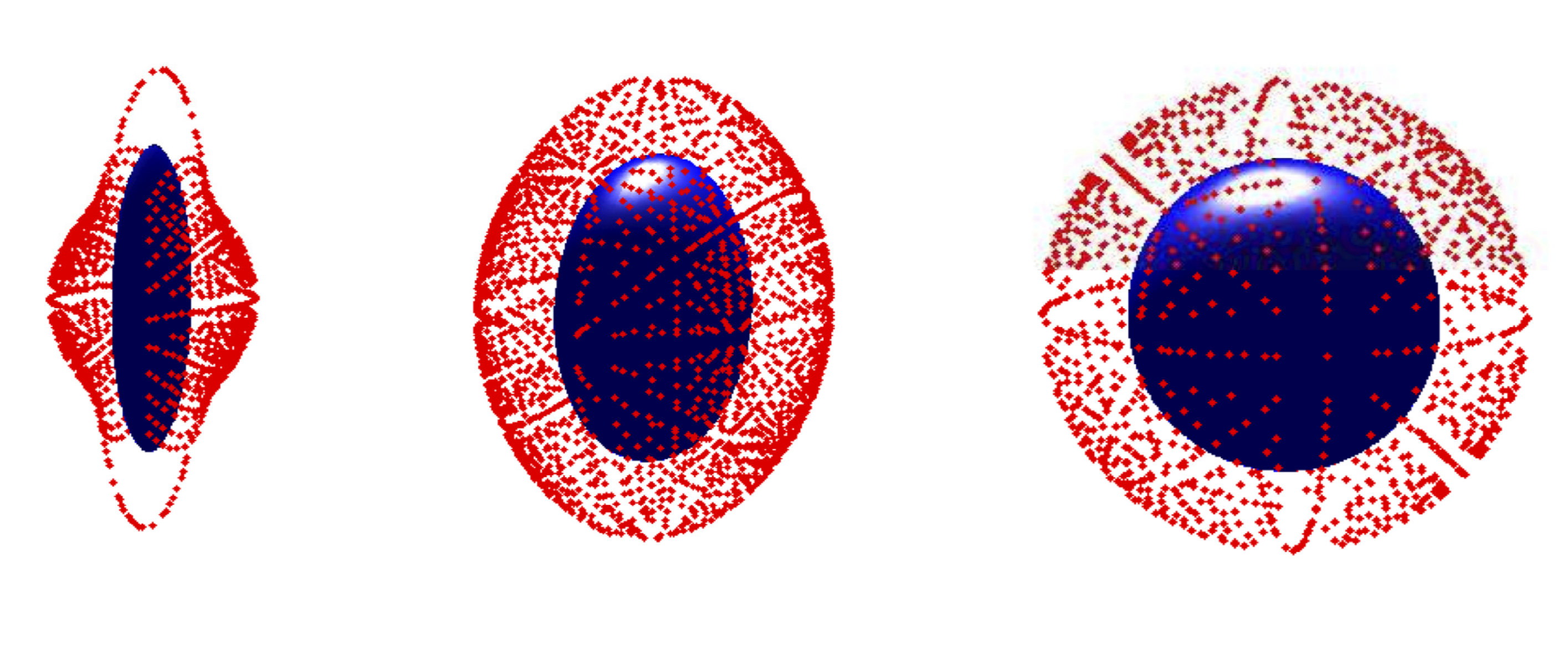}
\caption{ Plot of quasi-square-well potential under the condition $\mathbf{u}_i = \mathbf{u}_j$  for $\epsilon = $ 0.1, 0.5 and 0.9 from left to right. The red and blue surfaces represent the attraction range and hardcore repulsion associated with each ellipsoid, respectively.  View angle of each figure is tuned for better visualization. }
\label{fig:sigmaij}
\end{figure}

Analogous to  the patchy spherical model introduced by Kern and Frenkel \cite{Kern2003}, we choose the  ellipsoids to interact through a pair-potential that depends on their separation and orientation: $U_{ij} = U f(\mathbf{r}_{ij}, \mathbf{u}_i,\mathbf{u}_j)$. As illustrated for two oblate ellipsoids in Fig. \ref{fig:janusellipsoid},  attractive patches are coded by red and the orientation is chosen to coincide with the principle axis $ \mathbf{u}$ in the body-fixed frame of reference . 
The orientational interaction is defined as:
\begin{eqnarray}
f(\mathbf{r}_{ij}, \mathbf{u}_i,\mathbf{u}_j) =
 \begin{cases}
\ 1&  \mbox{if }\mathbf{u}_i \cdot \hat{\mathbf{r}}_{ij} \le \cos \delta , \mathbf{u}_j \cdot \hat{\mathbf{r}}_{ij} \ge -\cos \delta \\
\ 0& \mbox{otherwise}
\end{cases}
\end{eqnarray}
as shown in Fig. \ref{fig:janusellipsoid}. $\delta = \frac{\pi}{2}$ corresponds to Janus particle. An attractive interaction exists if two red patches face each other. 
The standard square-well potential has been used for Janus spheres; however,  the determination of the accurate spatial relation between ellipsoids is computationally time-consuming. 
We thus introduce a quasi-square-well potential defined as
\begin{eqnarray}
U=
\begin{cases}
 \infty& \mbox{if particles overlap}\\
 \ -U_0  H(\sigma_{ij}+0.5\sigma - r_{ij} )& \mbox{otherwise}\\
\end{cases}
\label{GeneralFreeEnergy}
\end{eqnarray}
 where $U_0$ is the well depth, $H(x)$ denotes the Heaviside function, $\sigma$ represents the length of the longer axis (max(2a,2b)). $r_{ij}$ is the center-to-center distance between ellipsoids: $r_{ij} = |\mathbf{r}_{ij}|$ with $\mathbf{r}_{ij} = \mathbf{r}_i -\mathbf{r}_j$.  
\begin{eqnarray}
\sigma_{ij}= 2b[ 1-\frac{\chi}{2}(\frac{(\hat{\mathbf{r}}_{ij} \cdot \mathbf{u}_i+\hat{\mathbf{r}}_{ij} \cdot \mathbf{u}_j)^2}{1+\chi  \mathbf{u}_i \cdot  \mathbf{u}_j} \nonumber\\
+\frac{(\hat{\mathbf{r}}_{ij} \cdot \mathbf{u}_i-\hat{\mathbf{r}}_{ij} \cdot \mathbf{u}_j)^2}{1-\chi  \mathbf{u}_i \cdot  \mathbf{u}_j})]^{-1/2}
\end{eqnarray}
with $\chi = \frac{\epsilon^2 -1}{\epsilon^2 +1}$, $\hat{\mathbf{r}}_{ij} = \frac{\mathbf{r}_{ij}}{r_{ij}}$ \cite{Berne1971}. $\sigma_{ij}$ is introduced as an approximation to characterize the spatial relation,  such that  there is no overlapping interaction if $r_{ij} \ge \sigma_{ij} $, provided that the ellipsoids are represented by a Gaussian function: $\exp(-\mathbf{r}\cdot \mathbf{\gamma}^{-1}\cdot \mathbf{r} )$ with $\mathbf{\gamma} = a^2 \mathbf{u}\mathbf{u} + b^2 (\mathbf{v}\mathbf{v}+\mathbf{w}\mathbf{w}) $  \cite{Berne1971}.  This approximation has been widely used to study anisotropic particles such as liquid crystals and granular material \cite{Care2005}. We note that in our study, this approximation is only applied to the potential but not to the geometric overlapping which will be determined using a precise method. Therefore, in our case $H(\sigma_{ij}+0.5\sigma - r_{ij} )$ represents a quasi-square-well potential with width $0.5\sigma$. Specific cases under the condition $\mathbf{u}_i = \mathbf{u}_j$ are illustrated in Fig. \ref{fig:sigmaij} for the aspect ratio $\epsilon$ = 0.1, 0.5 and 0.9 from left to right.  Two particles will interact if their red shells  touch each other.
The hard-core repulsion is provided by the following geometric relation. Given two ellipsoids $\mathcal{A}$: $\mathbf{X}^T\mathbf{A}\mathbf{X} = 0$ and $\mathcal{B}$: $\mathbf{X}^T\mathbf{B}\mathbf{X} = 0$, one introduces the characteristic polynomial $F(\lambda) = \det(\mathbf{A} - \lambda \mathbf{B})$. $\mathbf{A}$ and $\mathbf{B}$ are normalized so that the interiors of $\mathcal{A}$ and $\mathcal{B}$ satisfy $\mathbf{X}^T\mathbf{A}\mathbf{X} < 0$ and $\mathbf{X}^T\mathbf{B}\mathbf{X} < 0$. 
The roots of the characteristic equation $F(\lambda)  = 0$ have two positive real values and the rest characterizes the geometric relation between ellipsoids  \cite{Wang2001, Choi2009} such that \\
1. $\mathbf{A}$ and $\mathbf{B}$ are separate if and only if $F(\lambda) =0$ has two distinct negative roots; \\
2. $\mathbf{A}$ and $\mathbf{B}$ touch each other externally if and only if $F(\lambda) =0$ has a negative double root.\\
3. Otherwise, $\mathbf{A}$ and $\mathbf{B}$ overlap.\\
Sturm sequence  methods are applied to numerically decide if two roots are distinguishable. The primitive model we propose recovers the Kern-Frenkel Janus sphere model when $\epsilon = 1$,  since under this condition, $\sigma_{ij} = \sigma$ and consequently, the attraction is simplified to $H(1.5\sigma - r_{ij})$.
The introduced quasi-square-well potential has an advantage that there is no ambiguity when defining the connectivity of aggregates during the self-assembly process, and could be easily generalized to more realistic models. 

\begin{figure}[htbp]
\centering
\includegraphics[width=0.95\linewidth]{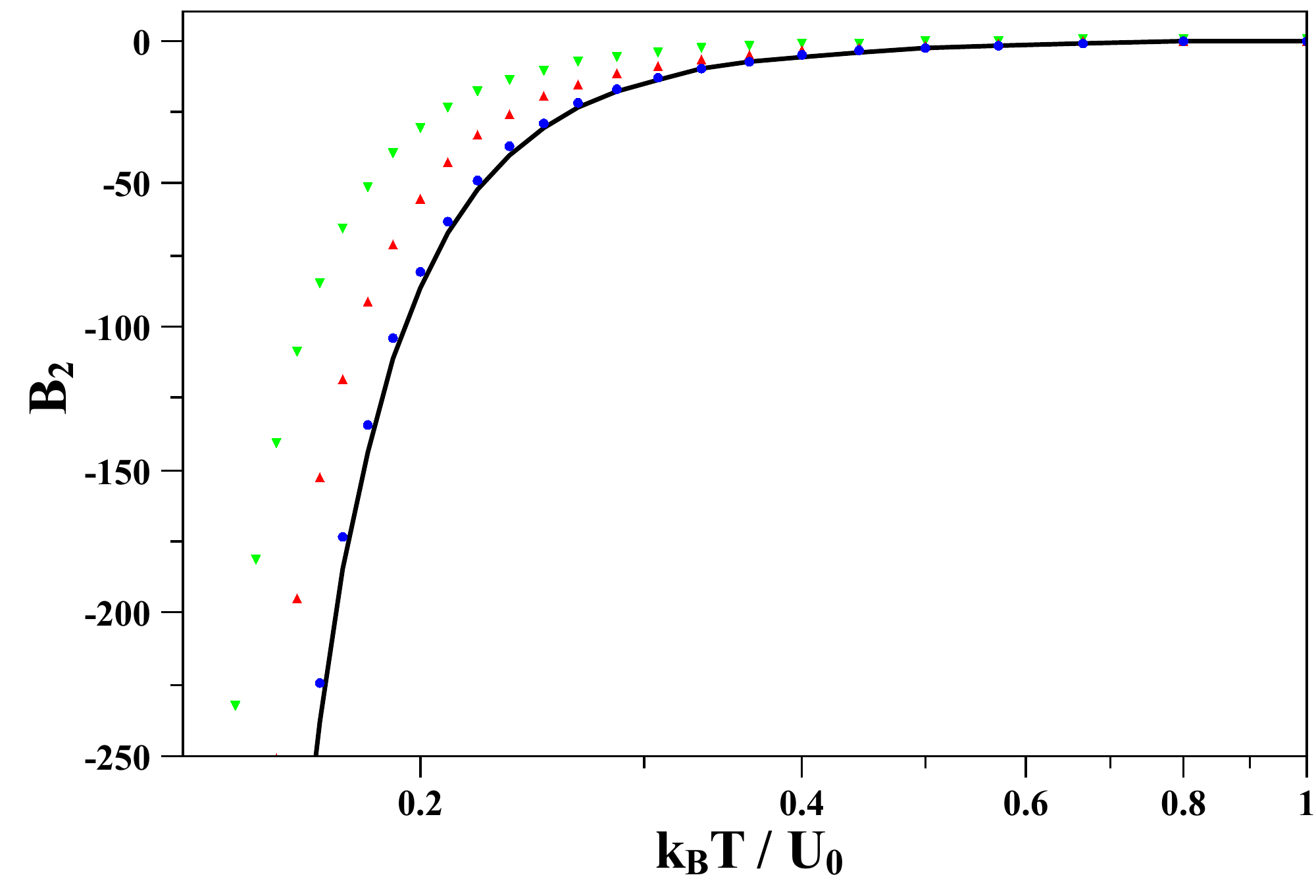}
\caption{ Plot of $B_2$ vs. $k_BT/U_0$ for the aspect ratio $\epsilon$ = 0.1 (green down triangle), 0.5 (red triangle) and 0.9 (blue circle) from top to bottom.  The solid curve is the theoretical prediction for a Janus sphere, i.e. $\epsilon = 1$.}
\label{fig:B2}
\end{figure}

Our model is  different from the Kern-Frenkel model when $\epsilon \neq 1$ since then the magnitude $U$ is a function of both the separation and orientation due to the anisotropic ellipsoidal core. The aspect ratio $\epsilon$ affects the shape of the interacting potentials.  To characterize the effective interaction between particles,  we calculate the second viral coefficient $B_2$:
\begin{eqnarray}
B_2= \frac{1}{2V}\frac{1}{(4\pi)^2}\int [1-e^{-\beta U_{12}}]\, d\mathbf{r}_1 d\mathbf{r}_2 d\mathbf{u}_1 d\mathbf{u}_2
\label{eq:b2}
\end{eqnarray}
The results of Monte Carlo integrations of $B_2/B_2^{hs}$ as a function of temperature for different $\epsilon$ are shown in Fig. \ref{fig:B2},  where without specification, $\epsilon = 0.1, 0.5$ and 0.9 are denoted by green down triangles, red triangles and blue circles, respectively.
Here $B_2^{hs} = \frac{2}{3}\pi \sigma^3$ stands for the second viral coefficient for hard spheres with diameter equal to $\sigma$. The numerical error is less than 1$\%$. The solid curve represents the theoretical prediction for a Janus sphere: $B_2/B_2^{hs} = 1-\frac{1}{4}(\delta^3 -1)(e^{\beta U_0} -1)$ with the interaction range $\delta = 1.5 \sigma$ in the study \cite{Kern2003}.  As $\epsilon$ increases to 1, $B_2$ approaches  the value for the Janus sphere from above. 

\section{Simulation results}
\begin{figure}[htbp]
\centering
\includegraphics[width=0.95\linewidth]{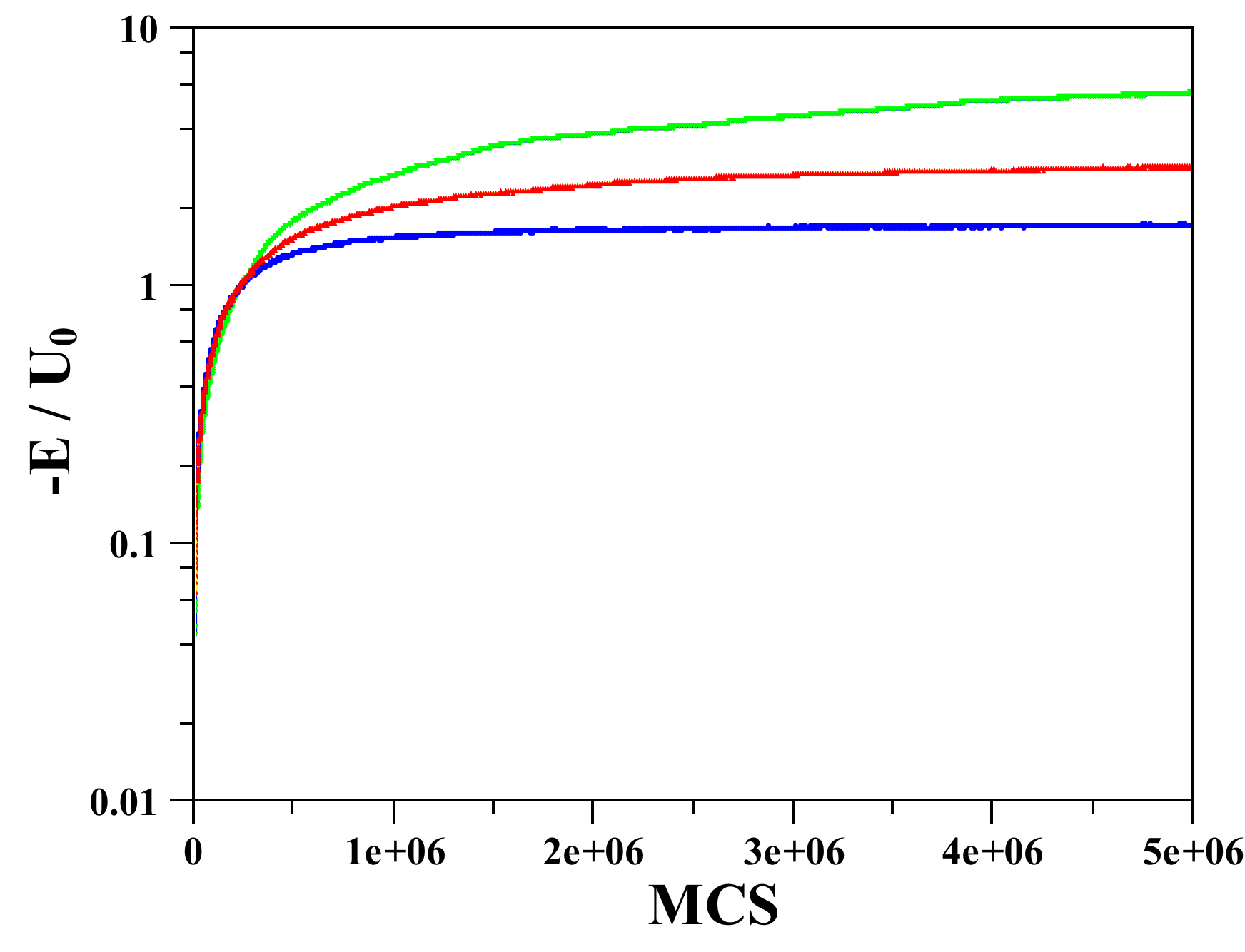}
\caption{ Plot of $-E/U_0$ vs. MCS. The curves from top to bottom correspond to the aspect ratio $\epsilon$ = 0.1 (green), 0.5 (red), and 0.9 (blue) }
\label{fig:Potential}
\end{figure}

Standard Monte Carlo (MC) simulation in the NVT ensemble has been applied to study the self-assembly pathway of these interacting Janus particles.  We investigate a system in a $30 \times 30 \times 30$ box with  periodic boundary conditions.  The particle aspect ratios range from 0.1 to 0.9  with number density $\rho = 0.037$;  the attractive energy is set as $\beta U_0 = 3$ with $\beta = \frac{1}{k_BT} $, where  $k_B$ is the Boltzmann constant. The system is initialized as a randomly-distributed noninteracting gas of monomers.  A random translation followed by a random rotation is carried out for each monomer. In particular, the rotation is performed using the method of quaternion parameters. More than ten independent runs for each aspect ratio have been carried out up to $5 \times 10^6$ Monte Carlo steps (MCS) and an ensemble average is taken by averaging over all runs.  We monitor the evolution of $-E/U_0$ (shown in Fig. \ref{fig:Potential}), the negative average potential scaled by the attraction strength, which characterizes the number of interacting neighbors of each particle. Since the initial system has no interaction, $-E/U_0$ starts from 0. It grows quickly and the dynamics slows down while approaching equilibrium. The system for $\epsilon = 0.9$ has reached equilibrium; however, for $\epsilon = 0.1$ and $0.5$ the system is only close to the equilibrium state after $5\times10^6$ MCS. This different growth tendency for the different aspect ratios is due to the distinct aggregation mechanism that dominates at different times.  As we will show in the later section, the aggregation at the early stage is dominated by  monomer diffusion and interactions with small,  formed oligomers.  Depending on the aspect ratio, the system is composed of monomers, small oligomers, micelles and vesicles. Afterwards, the aggregation dynamics is mainly the diffusion and collision of those small clusters that have much smaller diffusion constants than monomers. The value of $\epsilon$ affects the dynamics, as shown in Fig. \ref{fig:Potential}, such that the aggregation process is relatively faster but reaches a less stable structure when $\epsilon$ is larger. 

\begin{figure*}[htbp]
\centering
\includegraphics[width=0.95\linewidth]{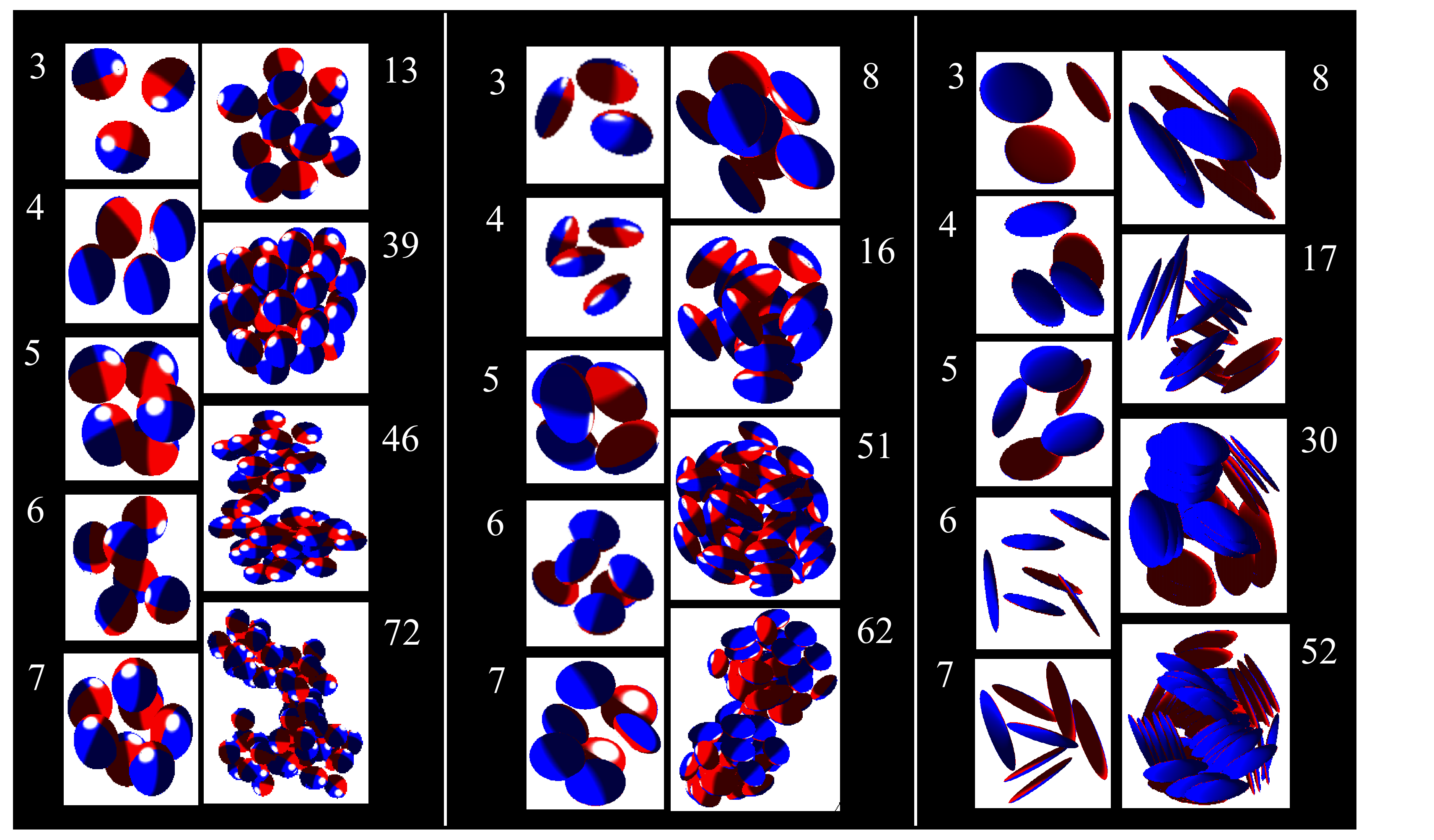}
\caption{ Plot of cluster formation for $\epsilon = 0.9, 0.5$ and $0.1$ from left to right. The number next to each panel is the cluster size. The size and view angle of each figure have been tuned for better visualization.}
\label{fig:oligomer}
\end{figure*}
To illustrate the difference in aggregation morphology, we take snapshots of the cluster growth as shown in Fig. \ref{fig:oligomer}. The cluster is defined such that two monomers are connected if they interact;  there is no ambiguity with respect to this interaction in our model. From left to right in Fig. \ref{fig:oligomer}, we show typical structures for $\epsilon$ = 0.1, 0.5, and 0.9 that are collected from all simulations. Small oligomers such as trimer, tetramer, pentamer, hexamer, and heptamer are similar and monomers form usual polygons, except that for $\epsilon = 0.1$ there is more space enclosed by monomers and the heptamer has a different structure: a rough double-layer. When the aspect ratio is less than 0.5, a different feature shows up. Instead of building up polygons, two ellipsoids with the same orientation pack into two layers. This is due to the quasi-square-well attraction in our model of the attraction range $0.5\sigma$. When $\epsilon < 0.5 $, two ellipsoids have  the possibility to interact  even when they are separated by the third one. As the oligomer grows, it starts forming a micelle (single layer) and a vesicle (double layer) as shown for clusters 13 and 39 ($\epsilon = 0.9$), and 8 and 16 ($\epsilon = 0.5$).  For $\epsilon = 0.1$, the relevant clusters (8 and 17) display two and three layers, respectively.  Those structures have been found in the case of Janus spheres and have an effect on breaking the thermal correlations between particles.  They thus affect the system phase behavior.  Eventually for $\epsilon = 0.9$, clusters with different structure are formed, including two small oligomers joining together (46), as well as more complicated chains (72).  For $\epsilon = 0.5$, it's possible to form a triple-layer compact cluster (51) and a dumbbell cluster(62).  For $\epsilon = 0.1$, due to the relation between the attractive range and aspect ratio, multiple-layer structures have been observed in the simulations (30, 52). 

\begin{figure}[htbp]
\centering
\includegraphics[width=0.95\linewidth]{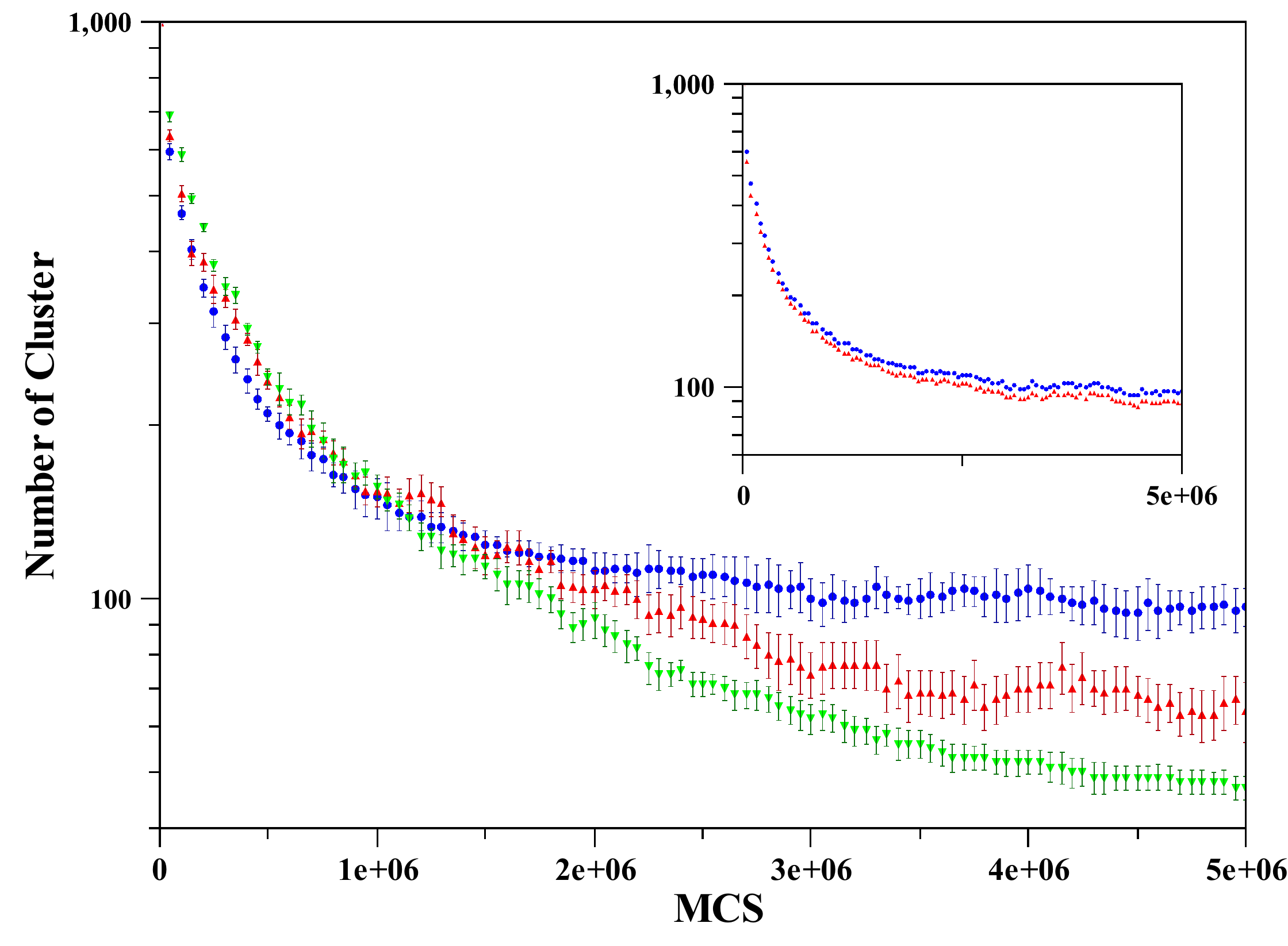}
\caption{ Plot of number of cluster vs. MCS for $\epsilon = $ 0.9 (blue circle), 0.5 (red up triangle) and 0.1 (green down triangle). Error bars come from the statistical variance of independent runs.  The inset shows the evolution of the number of clusters defined through energy (blue) and distance (red) for $\epsilon = 0.9$. }
\label{fig:Numcluster}
\end{figure}

As is well known, Monte Carlo simulation doesn't provide an accurate description of the kinetics of the system; however, it still reveals some useful features. We investigate the time evolution of number of clusters as illustrated in Fig. \ref{fig:Numcluster}.  The number of clusters for larger $\epsilon$ drops faster until reaching about $10^6$ MCS.  After that, the tendency reverses. The observation is consistent with the energy evolution and $B_2$. At the early stage, the dynamics is dominated by monomer motion and the system at $\epsilon = 0.9$ has a relatively stronger interaction which leads to faster cluster formation. Then the dynamics is governed by cluster-cluster interaction so that for larger aspect ratio the number of clusters decreases more slowly than for smaller aspect ratio.  

Note that the cluster defined by energy interaction might in principle be inconsistent with the experimental observations, which could implicitly use the distance definition without considering orientations. The inset in Fig. \ref{fig:Numcluster} shows that the difference between the energy definition (upper blue) and distance definition (lower red) for $\epsilon = 0.9$ is smaller than the statistical error, which indicates that the energy-defined cluster is a good approximation as compared with experiments.

\begin{figure}[htbp]
\centering
\includegraphics[width=1\linewidth]{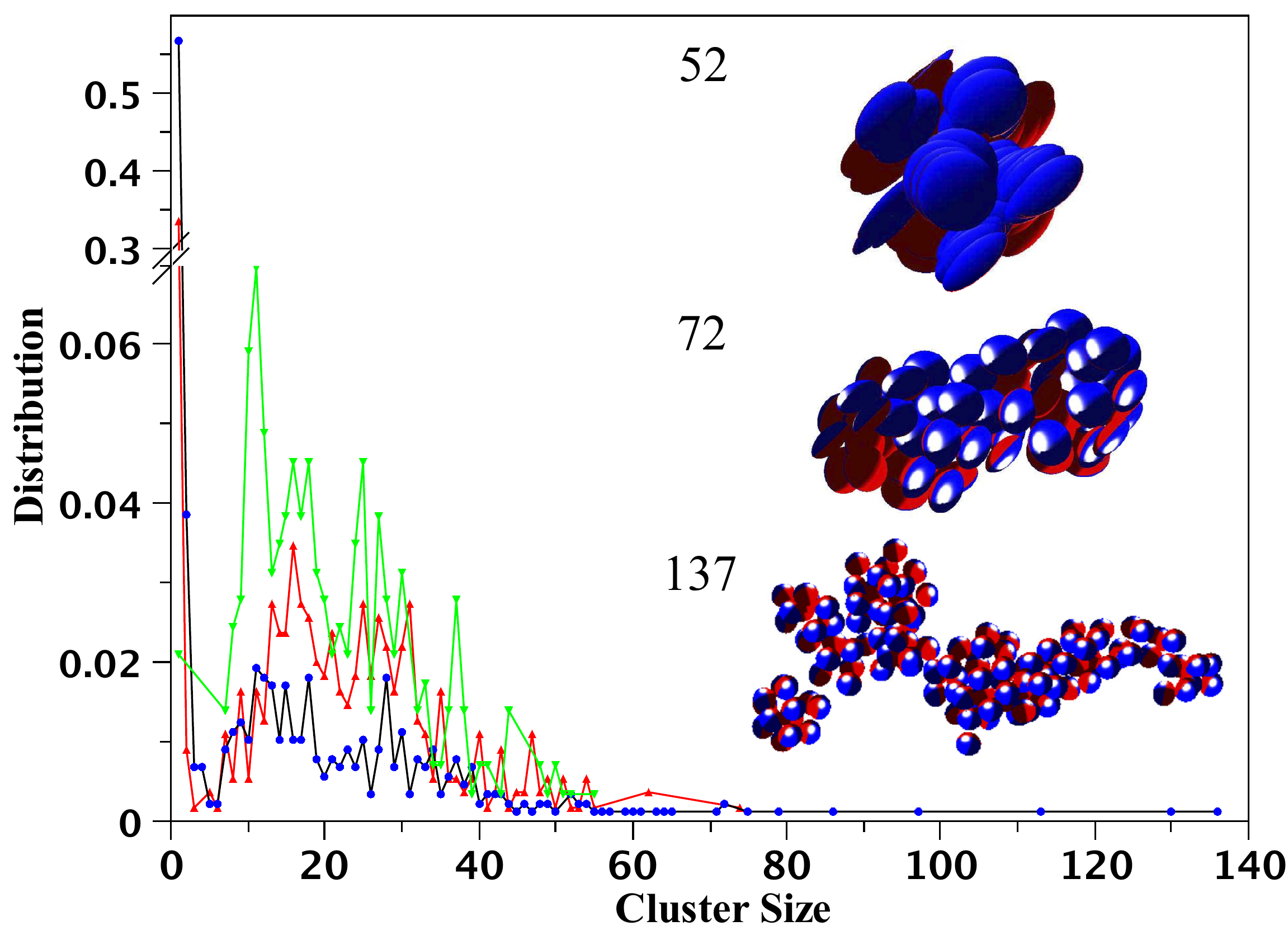}
\caption{ Plot of distribution of cluster size for $\epsilon =$ 0.1 (green down triangle), 0.5 (red triangle) and 0.9 (blue circle). Insets from top to bottom are configurations of the largest cluster found in the simulations for  $\epsilon$ = 0.1, 0.5 and 0.9, respectively. The number associated with each configuration is the cluster size.}
\label{fig:Clustersize}
\end{figure}

\begin{figure}[htbp]
\centering
\includegraphics[width=1\linewidth]{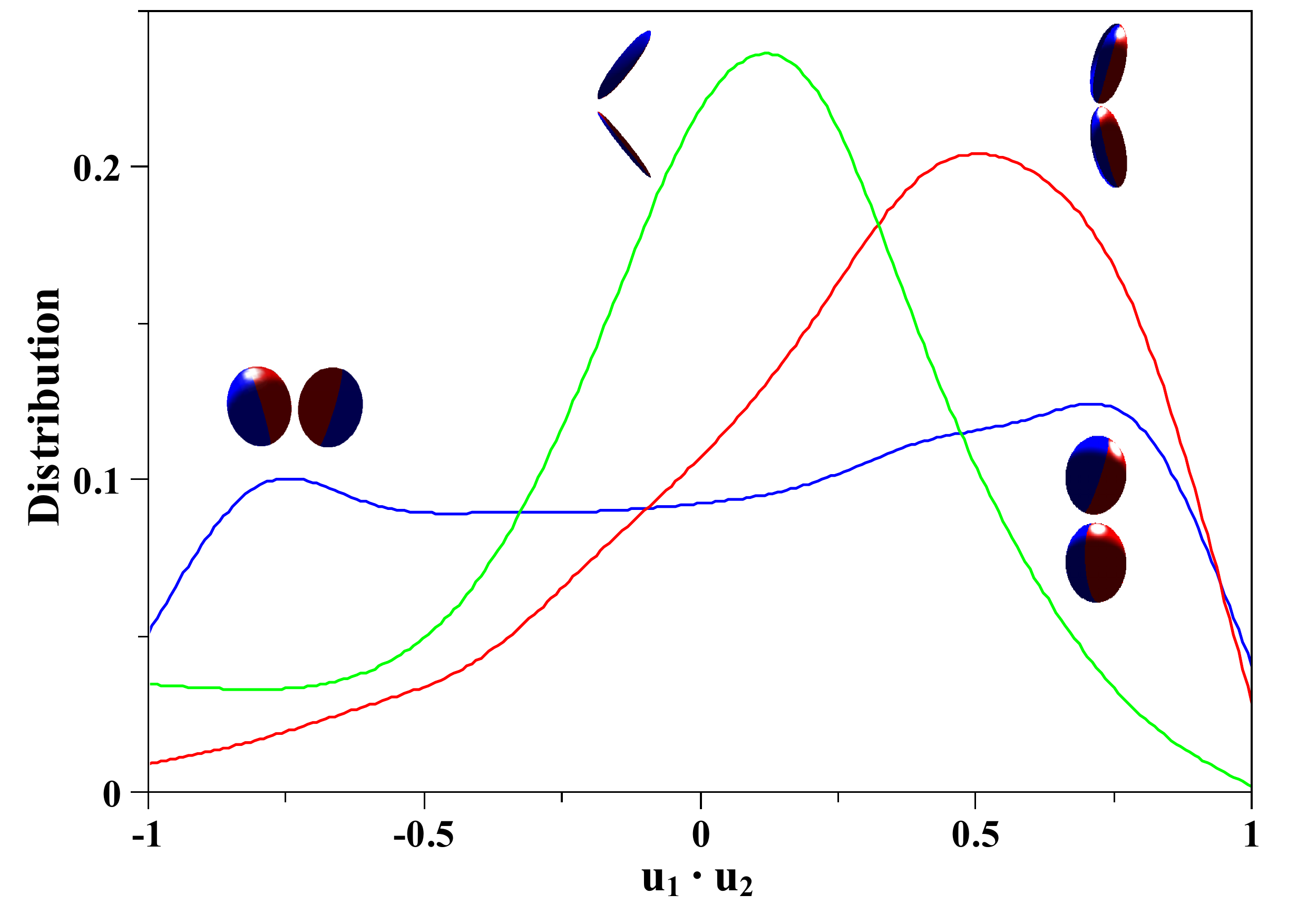}
\caption{ Plot of distribution of $\mathbf{u}_i \cdot \mathbf{u}_j$  between two bonded ellipsoids for $\epsilon =$ 0.1 (green), 0.5 (red) and 0.9 (blue).  Typical configurations corresponding to each peak are shown. }
\label{fig:Correlation}
\end{figure}

We have calculated the distribution of cluster size, which illustrates the effect of aspect ratio on the system approaching equilibrium, as shown in Fig. 
\ref{fig:Clustersize}. 
For larger $\epsilon$, the distribution has a broader range with a lower peak, which is consistent with the morphology shown in Fig. \ref{fig:oligomer}. Due to the  complex structure formation such as a chain, $\epsilon = 0.9$ has extended configurations with  larger size, in which the simulations show the largest cluster (a size 137 as shown in the inset of Fig.  \ref{fig:Clustersize}).  For $\epsilon = 0.1$, the distribution is narrow and has a higher peak, which indicates that the clusters are more uniform. The largest cluster with size 55 has a similar shape as the oligomer (52) shown in Fig. \ref{fig:oligomer}; the structure of this cluster is like a blob with several layers. The structure is more stable in terms  of its energy and prevents  further aggregation to form more complex forms.  The case of an aspect ratio 0.5 (74) is intermediate between these two extremes of 0.1 and 0.9  and the resulting cluster reveals two vesicles forming together through  a bridge-like structure. 
Next, in order to investigate in detail how ellipsoids organize in the cluster, we consider the correlation between patch orientations: $\mathbf{u}_i \cdot \mathbf{u}_j$ of two bonded ellipsoids (for which there exists a patchy attraction). The distribution $P(\mathbf{u}_i \cdot \mathbf{u}_j)$ is illustrated in Fig. \ref{fig:Correlation}. 

For $\epsilon = 0.9$, $P(\mathbf{u}_i, \mathbf{u}_j)$ develops two peaks; typical configurations corresponding to this correlation are shown in the figure. These two peaks correspond to
two ellipsoids facing toward ($\mathbf{u}_i \cdot \mathbf{u}_j \sim -1$) and opposite to ($\mathbf{u}_i \cdot \mathbf{u}_j\sim 1$) each other; similar behavior has been found for Janus spheres \cite{Sciortino2009}. As $\epsilon$ decreases, one peak ( $\mathbf{u}_i \cdot \mathbf{u}_j \sim -1$) vanishes, since ellipsoids self-assemble into a micelle-like structure (for example, panels 8 and 16 for $\epsilon = 0.5$ and panels 8 and 17 for $\epsilon = 0.1 $ in Fig. \ref{fig:oligomer}). Due to the geometric constraint, there is empty space inside the ellipsoids forming the cluster.  This property could be useful for the encapsulation of different types of particles (e.g. as in drug delivery). The other peak shifts towards zero; thus, the curvature is increasing so that the structure is more compact and energetically favorable. This structure prevents further aggregation, as the thermal fluctuation is suppressed, which prevents long range correlation. 

\begin{figure*}[htbp]
%\begin{wrapfigure}{l}
\centering
\includegraphics[width=1\linewidth]{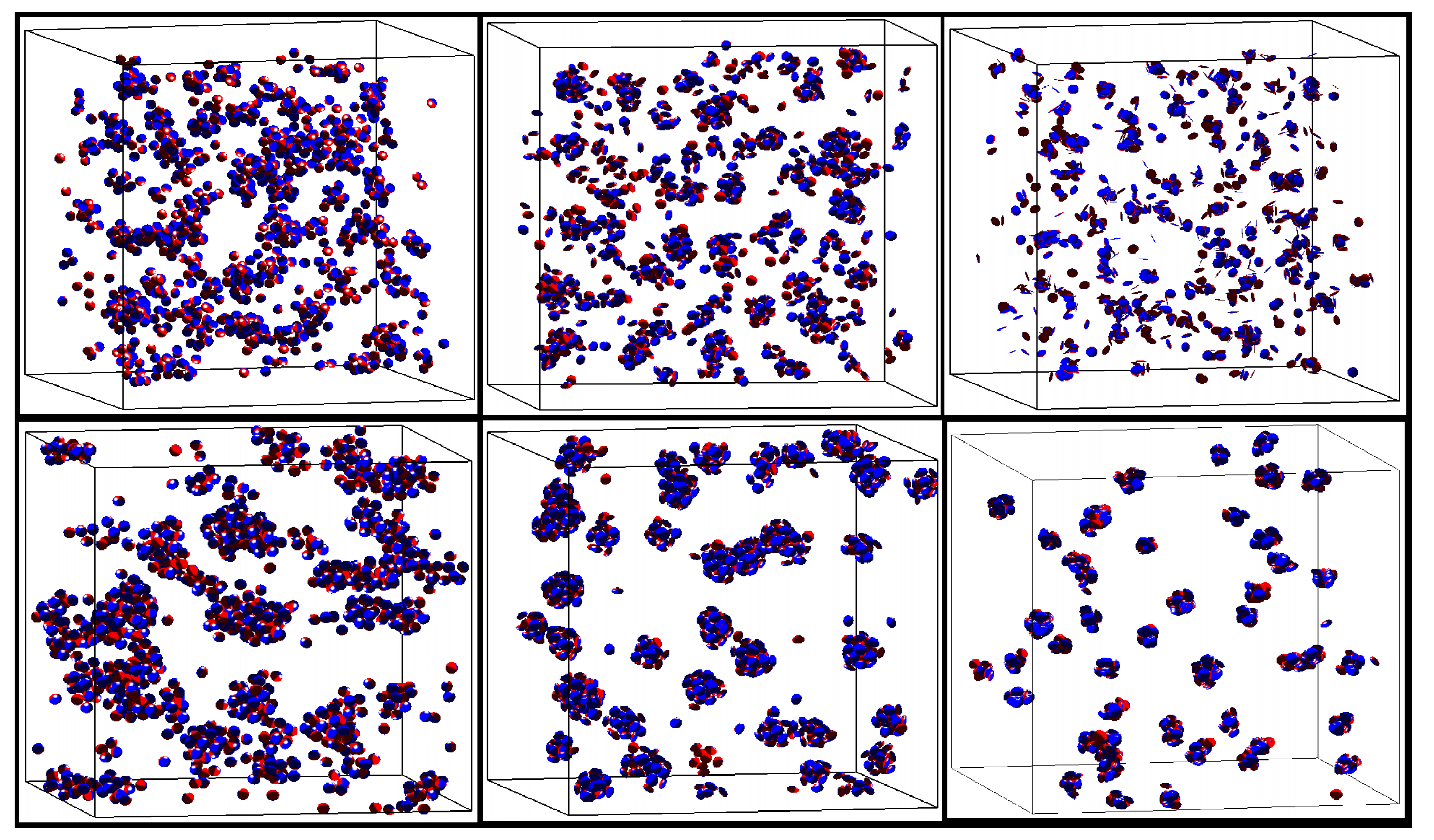}
\caption{ Snapshots of  the system for $\epsilon = 0.9, 0.5$ and 0.1 from left to right.  The upper and lower panels correspond to  $t = 10^6$ and $5 \times 10^6$ MCS.}
\label{fig:final}
%\end{wrapfigure}
\end{figure*}

More details could be explored from the snapshots of configurations.  As illustrated in Fig. \ref{fig:final}, the system for $\epsilon =0.9$ forms  large clusters at $10^6$ MCS and is composed of a large number of monomers, micelles, vesicles and large complex clusters at $5 \times 10^6$ MCS.

For $\epsilon = 0.5$, the configuration shows a somewhat different feature (middle panels of Fig. \ref{fig:final}) . The system forms micelles and vesicles at $10^6$ MCS and finally includes monomers, large numbers of micelles and vesicles, and their combined aggregates.  For $\epsilon = 0.1$, the configuration show that clusters forms slowly and only includes  small oligomers and a large number of micelles and vesicles at $5 \times 10^6$.

\section{Conclusion}
In the article, we propose a primitive model to study the self-assembly of Janus ellipsoids. The interactions between ellipsoids include a hard-core repulsion and a quasi-square-well attraction, where the latter exists when the patchy surfaces of two interacting ellipsoids orient in designated directions.  The anisotropy in our model comes from two aspects: the patch interaction and the anisotropic core, which are controlled by a patchy angle $\delta$ and an aspect ratio $\epsilon$, respectively. We particularly focus on the Janus ellipsoids in which $\delta = \frac{\pi}{2}$ and $\epsilon$ ranges from 0.1 to 0.9, and address the effects of aspect ratio on the self-assembly morphology and dynamical properties. Our model could be easily extended to consider more realistic systems. For example, the typical interaction between colloids is in the range  ($0.05\sigma \sim 0.2\sigma$), which however is computationally expensive to simulate. Janus ellipsoids with the interacting range comparable to $0.5\sigma$ can be realized by nanoparticles using the technique such as induced phase separation \cite{Shah2009}. Our results show that for larger aspect ratio, the self-assembly process is relatively faster and the morphology is more complicated, including chain structures as well as micelles and vesicles. The structures for smaller aspect ratio are more uniform and multiple-layer vesicles dominate.  In our Monte Carlo simulation, for $\epsilon = 0.1$ and 0.5, the system didn't reach equilibrium since when  multiple-layer vesicles form, the simulation is extremely time-consuming. However, the main features do not change, since the systems are very close to equilibrium. In our study, the phase diagram is not yet known, but our initial choice of parameters (such as density and temperature) puts us in the gas regime.  Our simple model raises many potential avenues for investigation; for example, the possible extension of $B_2$-scaling in the calculation of the phase diagram.  This $B_2$ scaling  has been successfully applied, for example, for patchy spheres \cite{Liu2009}. In addition, it is important see how our results depend on the range of the interaction, e.g. 0.2$\sigma$, as is the case for colloidal interactions.
We also find for small aspect ratio that ellipsoids tend to form vesicles, which suggests a potential application for particle encapsulation. Perhaps the most important result of the study for materials engineering is the fact that the size and structure of the aggregates can be controlled by the aspect ratio, which should be an interesting result from a design viewpoint.

\section{Acknowledgements}
This work was supported by grants from the Mathers Foundation and the National Science Foundation (Grant DMR-0702890).  One of us (GB) was supported by the NSF REU Site Grant in Physics at Lehigh University. Simulation work was supported in part by the National Science Foundation through TeraGrid resources provided by Pittsburgh Supercomputing Center. We thank Wenping Wang at HongKong University for providing the ellipsoid code.

\bibliographystyle{apsrev}
\bibliography{Janus_Ellipsoid.bib}

\begin{thebibliography}{27}
\expandafter\ifx\csname natexlab\endcsname\relax\def\natexlab#1{#1}\fi
\expandafter\ifx\csname bibnamefont\endcsname\relax
  \def\bibnamefont#1{#1}\fi
\expandafter\ifx\csname bibfnamefont\endcsname\relax
  \def\bibfnamefont#1{#1}\fi
\expandafter\ifx\csname citenamefont\endcsname\relax
  \def\citenamefont#1{#1}\fi
\expandafter\ifx\csname url\endcsname\relax
  \def\url#1{\texttt{#1}}\fi
\expandafter\ifx\csname urlprefix\endcsname\relax\def\urlprefix{URL }\fi
\providecommand{\bibinfo}[2]{#2}
\providecommand{\eprint}[2][]{\url{#2}}

\bibitem[{\citenamefont{Glotzer and Solomon}(2007)}]{Glotzer2007}
\bibinfo{author}{\bibfnamefont{S.~C.} \bibnamefont{Glotzer}} \bibnamefont{and}
  \bibinfo{author}{\bibfnamefont{M.~J.} \bibnamefont{Solomon}},
  \bibinfo{journal}{Nature Materials} \textbf{\bibinfo{volume}{6}},
  \bibinfo{pages}{557} (\bibinfo{year}{2007}).

\bibitem[{\citenamefont{Kegel and n.~w. Lekkerkerker}(2011)}]{Kegel2011}
\bibinfo{author}{\bibfnamefont{W.~K.} \bibnamefont{Kegel}} \bibnamefont{and}
  \bibinfo{author}{\bibfnamefont{H.}~\bibnamefont{n.~w. Lekkerkerker}},
  \bibinfo{journal}{Nature Materials} \textbf{\bibinfo{volume}{10}},
  \bibinfo{pages}{5} (\bibinfo{year}{2011}).

\bibitem[{\citenamefont{Chen et~al.}(2011{\natexlab{a}})\citenamefont{Chen,
  K.Whitmer, Jiang, Bae, Luijten, and Granick}}]{QianChen2011a}
\bibinfo{author}{\bibfnamefont{Q.}~\bibnamefont{Chen}},
  \bibinfo{author}{\bibfnamefont{J.}~\bibnamefont{K.Whitmer}},
  \bibinfo{author}{\bibfnamefont{S.}~\bibnamefont{Jiang}},
  \bibinfo{author}{\bibfnamefont{S.~C.} \bibnamefont{Bae}},
  \bibinfo{author}{\bibfnamefont{E.}~\bibnamefont{Luijten}}, \bibnamefont{and}
  \bibinfo{author}{\bibfnamefont{S.}~\bibnamefont{Granick}},
  \bibinfo{journal}{Science} \textbf{\bibinfo{volume}{331}},
  \bibinfo{pages}{199} (\bibinfo{year}{2011}{\natexlab{a}}).

\bibitem[{\citenamefont{Pawar and Kretzschmar}(2009)}]{Pawar2009}
\bibinfo{author}{\bibfnamefont{A.~B.} \bibnamefont{Pawar}} \bibnamefont{and}
  \bibinfo{author}{\bibfnamefont{I.}~\bibnamefont{Kretzschmar}},
  \bibinfo{journal}{Langmuir} \textbf{\bibinfo{volume}{25}},
  \bibinfo{pages}{9057} (\bibinfo{year}{2009}).

\bibitem[{\citenamefont{Chen et~al.}(2011{\natexlab{b}})\citenamefont{Chen,
  Bae, and Granick}}]{QianChen2011}
\bibinfo{author}{\bibfnamefont{Q.}~\bibnamefont{Chen}},
  \bibinfo{author}{\bibfnamefont{S.~C.} \bibnamefont{Bae}}, \bibnamefont{and}
  \bibinfo{author}{\bibfnamefont{S.}~\bibnamefont{Granick}},
  \bibinfo{journal}{Nature} \textbf{\bibinfo{volume}{469}},
  \bibinfo{pages}{381} (\bibinfo{year}{2011}{\natexlab{b}}).

\bibitem[{\citenamefont{Romano and Sciortino}(2011)}]{Romano2011}
\bibinfo{author}{\bibfnamefont{F.}~\bibnamefont{Romano}} \bibnamefont{and}
  \bibinfo{author}{\bibfnamefont{F.}~\bibnamefont{Sciortino}},
  \bibinfo{journal}{Nature Materials} \textbf{\bibinfo{volume}{10}},
  \bibinfo{pages}{171} (\bibinfo{year}{2011}).

\bibitem[{\citenamefont{Pawar and Kretzschmar}(2010)}]{Pawar2010}
\bibinfo{author}{\bibfnamefont{A.~B.} \bibnamefont{Pawar}} \bibnamefont{and}
  \bibinfo{author}{\bibfnamefont{I.}~\bibnamefont{Kretzschmar}},
  \bibinfo{journal}{Macromol. Rapid Commun.} \textbf{\bibinfo{volume}{32}},
  \bibinfo{pages}{150} (\bibinfo{year}{2010}).

\bibitem[{\citenamefont{Jiang et~al.}(2010)\citenamefont{Jiang, Chen, Tripathy,
  Luijten, Schweizer, and Granick}}]{Jiang2010}
\bibinfo{author}{\bibfnamefont{S.}~\bibnamefont{Jiang}},
  \bibinfo{author}{\bibfnamefont{Q.}~\bibnamefont{Chen}},
  \bibinfo{author}{\bibfnamefont{M.}~\bibnamefont{Tripathy}},
  \bibinfo{author}{\bibfnamefont{E.}~\bibnamefont{Luijten}},
  \bibinfo{author}{\bibfnamefont{K.~S.} \bibnamefont{Schweizer}},
  \bibnamefont{and} \bibinfo{author}{\bibfnamefont{S.}~\bibnamefont{Granick}},
  \bibinfo{journal}{Adv. Mater.} \textbf{\bibinfo{volume}{22}},
  \bibinfo{pages}{1060} (\bibinfo{year}{2010}).

\bibitem[{\citenamefont{Manoharan et~al.}(2003)\citenamefont{Manoharan,
  M.T.Elsesser, and D.J.Pine}}]{Manoharan2003}
\bibinfo{author}{\bibfnamefont{V.~N.} \bibnamefont{Manoharan}},
  \bibinfo{author}{\bibnamefont{M.T.Elsesser}}, \bibnamefont{and}
  \bibinfo{author}{\bibnamefont{D.J.Pine}}, \bibinfo{journal}{Science}
  \textbf{\bibinfo{volume}{301}}, \bibinfo{pages}{483} (\bibinfo{year}{2003}).

\bibitem[{\citenamefont{Zhang and Glotzer}(2004)}]{Zhang2004}
\bibinfo{author}{\bibfnamefont{Z.}~\bibnamefont{Zhang}} \bibnamefont{and}
  \bibinfo{author}{\bibfnamefont{S.~C.} \bibnamefont{Glotzer}},
  \bibinfo{journal}{Nano Lett.} \textbf{\bibinfo{volume}{4}},
  \bibinfo{pages}{1407} (\bibinfo{year}{2004}).

\bibitem[{\citenamefont{Hong et~al.}(2006)\citenamefont{Hong, Jiang, and
  Granick}}]{Hong2006}
\bibinfo{author}{\bibfnamefont{L.}~\bibnamefont{Hong}},
  \bibinfo{author}{\bibfnamefont{S.}~\bibnamefont{Jiang}}, \bibnamefont{and}
  \bibinfo{author}{\bibfnamefont{S.}~\bibnamefont{Granick}},
  \bibinfo{journal}{Langmuir} \textbf{\bibinfo{volume}{22}},
  \bibinfo{pages}{9495} (\bibinfo{year}{2006}).

\bibitem[{\citenamefont{Sciortino et~al.}(2009)\citenamefont{Sciortino,
  Giacometti, and Pastore}}]{Sciortino2009}
\bibinfo{author}{\bibfnamefont{F.}~\bibnamefont{Sciortino}},
  \bibinfo{author}{\bibfnamefont{A.}~\bibnamefont{Giacometti}},
  \bibnamefont{and} \bibinfo{author}{\bibfnamefont{G.}~\bibnamefont{Pastore}},
  \bibinfo{journal}{Phys. Rev. Lett.} \textbf{\bibinfo{volume}{103}},
  \bibinfo{pages}{237801} (\bibinfo{year}{2009}).

\bibitem[{\citenamefont{Miller and Cacciuto}(2009)}]{Miller2009}
\bibinfo{author}{\bibfnamefont{W.~L.} \bibnamefont{Miller}} \bibnamefont{and}
  \bibinfo{author}{\bibfnamefont{A.}~\bibnamefont{Cacciuto}},
  \bibinfo{journal}{Phys. Rev. E} \textbf{\bibinfo{volume}{80}},
  \bibinfo{pages}{021404} (\bibinfo{year}{2009}).

\bibitem[{\citenamefont{Sciortino et~al.}(2010)\citenamefont{Sciortino,
  Giacometti, and Pastore}}]{Sciortino2010}
\bibinfo{author}{\bibfnamefont{F.}~\bibnamefont{Sciortino}},
  \bibinfo{author}{\bibfnamefont{A.}~\bibnamefont{Giacometti}},
  \bibnamefont{and} \bibinfo{author}{\bibfnamefont{G.}~\bibnamefont{Pastore}},
  \bibinfo{journal}{Phys. Chem. Chem. Phys.} \textbf{\bibinfo{volume}{12}},
  \bibinfo{pages}{11869} (\bibinfo{year}{2010}).

\bibitem[{\citenamefont{Kern and Frenkel}(2003)}]{Kern2003}
\bibinfo{author}{\bibfnamefont{N.}~\bibnamefont{Kern}} \bibnamefont{and}
  \bibinfo{author}{\bibfnamefont{D.}~\bibnamefont{Frenkel}},
  \bibinfo{journal}{J. Chem. Phys.} \textbf{\bibinfo{volume}{118}},
  \bibinfo{pages}{9882} (\bibinfo{year}{2003}).

\bibitem[{\citenamefont{Reinhardt et~al.}(2011)\citenamefont{Reinhardt,
  Williamson, Doye, Carrete, Varela, and Louis}}]{Reinhardt2011}
\bibinfo{author}{\bibfnamefont{A.}~\bibnamefont{Reinhardt}},
  \bibinfo{author}{\bibfnamefont{A.~J.} \bibnamefont{Williamson}},
  \bibinfo{author}{\bibfnamefont{J.~P.~K.} \bibnamefont{Doye}},
  \bibinfo{author}{\bibfnamefont{J.}~\bibnamefont{Carrete}},
  \bibinfo{author}{\bibfnamefont{L.~M.} \bibnamefont{Varela}},
  \bibnamefont{and} \bibinfo{author}{\bibfnamefont{A.~A.} \bibnamefont{Louis}},
  \bibinfo{journal}{J. Chem. Phys.} \textbf{\bibinfo{volume}{134}},
  \bibinfo{pages}{104905} (\bibinfo{year}{2011}).

\bibitem[{\citenamefont{Fejer and Wales}(2007)}]{Fejer2007}
\bibinfo{author}{\bibfnamefont{S.~N.} \bibnamefont{Fejer}} \bibnamefont{and}
  \bibinfo{author}{\bibfnamefont{D.~J.} \bibnamefont{Wales}},
  \bibinfo{journal}{Phys. Rev. Lett.} \textbf{\bibinfo{volume}{99}},
  \bibinfo{pages}{086106} (\bibinfo{year}{2007}).

\bibitem[{\citenamefont{Fejer et~al.}(2011)\citenamefont{Fejer, Chakrabraty,
  and Wales}}]{Fejer2011}
\bibinfo{author}{\bibfnamefont{S.~N.} \bibnamefont{Fejer}},
  \bibinfo{author}{\bibfnamefont{D.}~\bibnamefont{Chakrabraty}},
  \bibnamefont{and} \bibinfo{author}{\bibfnamefont{D.~J.} \bibnamefont{Wales}},
  \bibinfo{journal}{Soft Matter} \textbf{\bibinfo{volume}{7}},
  \bibinfo{pages}{3553} (\bibinfo{year}{2011}).

\bibitem[{\citenamefont{van Dillen et~al.}(2004)\citenamefont{van Dillen, van
  Blaaderen, and Polman}}]{Dillen2004}
\bibinfo{author}{\bibfnamefont{T.}~\bibnamefont{van Dillen}},
  \bibinfo{author}{\bibfnamefont{A.}~\bibnamefont{van Blaaderen}},
  \bibnamefont{and} \bibinfo{author}{\bibfnamefont{A.}~\bibnamefont{Polman}},
  \bibinfo{journal}{Materials Today} \textbf{\bibinfo{volume}{7}},
  \bibinfo{pages}{40} (\bibinfo{year}{2004}).

\bibitem[{\citenamefont{Ho et~al.}(1993)\citenamefont{Ho, Keller, Odell, and
  Ottewill}}]{Ho1993}
\bibinfo{author}{\bibfnamefont{C.~C.} \bibnamefont{Ho}},
  \bibinfo{author}{\bibfnamefont{A.}~\bibnamefont{Keller}},
  \bibinfo{author}{\bibfnamefont{J.~A.} \bibnamefont{Odell}}, \bibnamefont{and}
  \bibinfo{author}{\bibfnamefont{R.~H.} \bibnamefont{Ottewill}},
  \bibinfo{journal}{Colliod and Polymer Science}
  \textbf{\bibinfo{volume}{271}}, \bibinfo{pages}{469} (\bibinfo{year}{1993}).

\bibitem[{\citenamefont{DeConinck et~al.}()\citenamefont{DeConinck, Shepherd,
  Cote, Granick, Schweizer, and Lewis}}]{DeConinck}
\bibinfo{author}{\bibfnamefont{A.~J.} \bibnamefont{DeConinck}},
  \bibinfo{author}{\bibfnamefont{R.~F.} \bibnamefont{Shepherd}},
  \bibinfo{author}{\bibfnamefont{A.~R.} \bibnamefont{Cote}},
  \bibinfo{author}{\bibfnamefont{S.}~\bibnamefont{Granick}},
  \bibinfo{author}{\bibfnamefont{K.~S.} \bibnamefont{Schweizer}},
  \bibnamefont{and} \bibinfo{author}{\bibfnamefont{J.~A.} \bibnamefont{Lewis}},
  \bibinfo{note}{poster (MRL 2009)}.

\bibitem[{\citenamefont{Berne and Pechukas}(1971)}]{Berne1971}
\bibinfo{author}{\bibfnamefont{B.~J.} \bibnamefont{Berne}} \bibnamefont{and}
  \bibinfo{author}{\bibfnamefont{P.}~\bibnamefont{Pechukas}},
  \bibinfo{journal}{J. Chem. Phys.} \textbf{\bibinfo{volume}{56}},
  \bibinfo{pages}{4213} (\bibinfo{year}{1971}).

\bibitem[{\citenamefont{Care and Cleaver}(2005)}]{Care2005}
\bibinfo{author}{\bibfnamefont{C.~M.} \bibnamefont{Care}} \bibnamefont{and}
  \bibinfo{author}{\bibfnamefont{D.~J.} \bibnamefont{Cleaver}},
  \bibinfo{journal}{Rep. Prog. Phys.} \textbf{\bibinfo{volume}{68}},
  \bibinfo{pages}{2665} (\bibinfo{year}{2005}).

\bibitem[{\citenamefont{Wang et~al.}(2001)\citenamefont{Wang, Wang, and
  Kim}}]{Wang2001}
\bibinfo{author}{\bibfnamefont{W.}~\bibnamefont{Wang}},
  \bibinfo{author}{\bibfnamefont{J.}~\bibnamefont{Wang}}, \bibnamefont{and}
  \bibinfo{author}{\bibfnamefont{M.-S.} \bibnamefont{Kim}},
  \bibinfo{journal}{Computer Aided Geometric Design}
  \textbf{\bibinfo{volume}{18}}, \bibinfo{pages}{531} (\bibinfo{year}{2001}).

\bibitem[{\citenamefont{Choi et~al.}(2009)\citenamefont{Choi, Chang, Wang, Kim,
  and Elber}}]{Choi2009}
\bibinfo{author}{\bibfnamefont{Y.-K.} \bibnamefont{Choi}},
  \bibinfo{author}{\bibfnamefont{J.-W.} \bibnamefont{Chang}},
  \bibinfo{author}{\bibfnamefont{W.}~\bibnamefont{Wang}},
  \bibinfo{author}{\bibfnamefont{M.-S.} \bibnamefont{Kim}}, \bibnamefont{and}
  \bibinfo{author}{\bibfnamefont{G.}~\bibnamefont{Elber}},
  \bibinfo{journal}{IEEE Trans.Visualization and Computer Graphics}
  \textbf{\bibinfo{volume}{15}}, \bibinfo{pages}{311} (\bibinfo{year}{2009}).

\bibitem[{\citenamefont{Shah et~al.}(2009)\citenamefont{Shah, Kim, and
  Weitz}}]{Shah2009}
\bibinfo{author}{\bibfnamefont{R.~K.} \bibnamefont{Shah}},
  \bibinfo{author}{\bibfnamefont{J.-W.} \bibnamefont{Kim}}, \bibnamefont{and}
  \bibinfo{author}{\bibfnamefont{D.~A.} \bibnamefont{Weitz}},
  \bibinfo{journal}{Adv. Mater.} \textbf{\bibinfo{volume}{21}},
  \bibinfo{pages}{1949} (\bibinfo{year}{2009}).

\bibitem[{\citenamefont{Liu et~al.}(2009)\citenamefont{Liu, Kumar, Sciortino,
  and Evans}}]{Liu2009}
\bibinfo{author}{\bibfnamefont{H.}~\bibnamefont{Liu}},
  \bibinfo{author}{\bibfnamefont{S.~K.} \bibnamefont{Kumar}},
  \bibinfo{author}{\bibfnamefont{F.}~\bibnamefont{Sciortino}},
  \bibnamefont{and} \bibinfo{author}{\bibfnamefont{G.~T.} \bibnamefont{Evans}},
  \bibinfo{journal}{J. Chem. Phys.} \textbf{\bibinfo{volume}{130}},
  \bibinfo{pages}{044902} (\bibinfo{year}{2009}).

\end{thebibliography}

\end{document}